# RAINBOW FORMATION: FROM DESCARTES TO VENUS

A.D. Zaikin

Novosibirsk State Technical University, Novosibirsk, Russia

The mechanism of rainbow formation proposed by Descartes and Newton is analyzed. The parameters of rainbows on Venus are calculated using geometric optics. Assuming that solar radiation is refracted by spherical droplets of aqueous sulfuric acid solution, the angular dimensions of the primary and secondary rainbows are determined as functions of the solution's concentration. It is shown that changes in concentration have the greatest effect on the size of the Alexander's dark band, the angular distance between the primary and secondary rainbows.

**Key words:** rainbow, Descartes, geometric optics, light dispersion, sulfuric acid, Venus.

## Introduction

A rainbow, observed when the Sun illuminates a curtain of rain, is a very common optical phenomenon. Along with aesthetic pleasure, it naturally raises questions about the causes of its appearance. The first scientific publication on the mechanism of rainbow formation, by Descartes, was published in 1637. Applying the laws of geometric optics, he explained [1] the formation of a rainbow by the refraction of sunlight in water droplets. Based on the phenomenon of light dispersion, Newton elucidated [2] the mechanism of formation of a colored rainbow. The wave theory of light, based on the work of Young and Airy, made it possible to describe the subtle effects of the formation of higher-order arcs in a rainbow. The exact solution, constructed by Mie, is based on the calculation of the scattering of an electromagnetic wave by a spherical inclusion. Review works [3,4] provide an understanding of the wave theory of the rainbow.

The existence of an exact numerical solution has not completely exhausted the topic of rainbows. Analyzing such solutions often proves more difficult than constructing them. Analysis of the physical nature of various optical phenomena observed in the atmosphere, similar to the classic rainbow, remains an active topic.

Eyewitnesses observing the Tunguska event noted the formation of a rainbow trail along its trajectory. According to the authors' point of view, as outlined in [5], the presence of a rainbow trail is evidence that this celestial body could only have been a comet, from which water was released directly during its flight in the mesopause.

A glory rainbow is a colored ring of light on clouds around the shadow of an observer positioned on a mountain or in an airplane. The light source (the Sun or Moon) is located behind the observer. While a rainbow is caused by light refraction, a glory rainbow is the result of light diffraction by very small, uniform droplets.

In [6], it was shown that glory is formed in clouds with negative temperatures on liquid spherical particles of amorphous water with a refractive index of 1.81-1.82 and a diameter of more than 20 μm. The geometric characteristics of glory allow for remotely obtaining information about dispersed phases in cold clouds.

The study of iridescent phenomena in extraterrestrial atmospheres deserves special mention. On July 24, 2011, ESA's VenusExpress orbiter detected a rainbow phenomenon (gloria) at an altitude of 70 km above the surface of Venus [7]. Observing the glory required special positioning of the research spacecraft; it had to be positioned directly between the Sun and a cloud reflecting sunlight. The phenomenon can be explained by light diffraction.

In this paper, we limit ourselves to the geometric optics approximation, which correctly describes the angular positions of rainbow extrema. We use this approach to analyze the parameters of a classical rainbow occurring when light is scattered by sulfuric acid droplets.

### The path of a light beam in a droplet upon single reflection

A beam of light falls on a spherical drop of radius $R$. The distance from the beam to a line parallel to it and passing through the center of the sphere is called the impact parameter. Denoting the impact parameter as $\rho$, we also define the dimensionless impact parameter $\delta=\rho/R$. The angle between the beam and the normal to the sphere's surface at the point of impact is called the angle of incidence and is denoted by $\alpha$. We assume that the source of the parallel rays (the Sun), the beam, the center of the drop, and the observer's eye are all in the same vertical plane. The observer, located between the Sun and the cloud of droplets, is positioned below them.

Consider a ray incident on the upper hemisphere (Fig. 1). Refracted at point *A* at an angle *β*, the ray reaches the surface of the sphere at point *B*, is reflected, and, refracted at point *C*, exits the sphere in the direction opposite to the incident ray. Reflection of the ray also occurs at points *A* and *C*, and refraction at point *B*; however, in this formulation, these phenomena are unimportant.

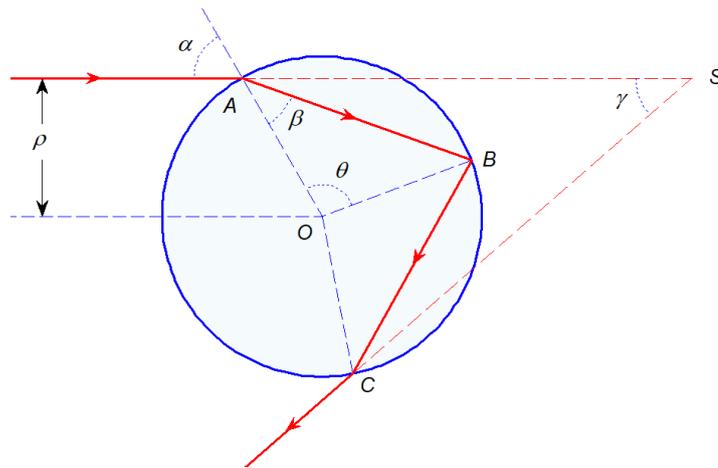

*Fig. 1. Single reflection of a beam in a spherical drop*

Since the light beam can also fall on the lower hemisphere, the values of the angle *α* lie in the range $\pm \pi/2$, and $-1 \leq \delta \leq 1$, since $\delta = \sin \alpha$.

The angle of incidence and the angle of reflection at point *B* are equal. Triangles *AOB* and *BOC* are isosceles, so all four base angles are equal to *β*. The angle formed by the radii in these triangles is denoted as *θ*.

Extending the rays incident on the sphere and emerging from it until they intersect at point *S*. The ray emerging from the drop will deviate from the direction of the light source by an angle *γ*. We will count it counterclockwise from the direction of the source.

For an isosceles triangle *ASC* it is true

$$AC^2 = AS^2 + SC^2 - 2AS \cdot SC \cdot \cos\gamma .$$

Since $AC = 2R\sin\theta$, then

$$AS = \frac{AC}{\sqrt{2(1-\cos\gamma)}} = \frac{\sqrt{2}R\sin\theta}{\sqrt{1-\cos\gamma}}.$$

For triangle $AOB$ $\theta + \beta + \beta = \pi$, and for triangle $AOS$ $\theta + \alpha + \gamma/2 = \pi$. Subtracting these relations, we obtain that

$$\gamma = 2(2\beta - \alpha).$$

Let $n$ be the refractive index of the droplet material (in our case, water). Let the refractive index of air be equal to one, then Snell's law implies that

$$\sin\alpha = n\sin\beta, \quad \beta = \arcsin(\sin\alpha/n),$$

and the angle of deflection of the beam is equal to

$$\gamma = 4\arcsin\left(\frac{\sin\alpha}{n}\right) - 2\alpha. \tag{1}$$

The graph of the obtained function $\gamma(\alpha, n)$ is shown in Fig. 2.

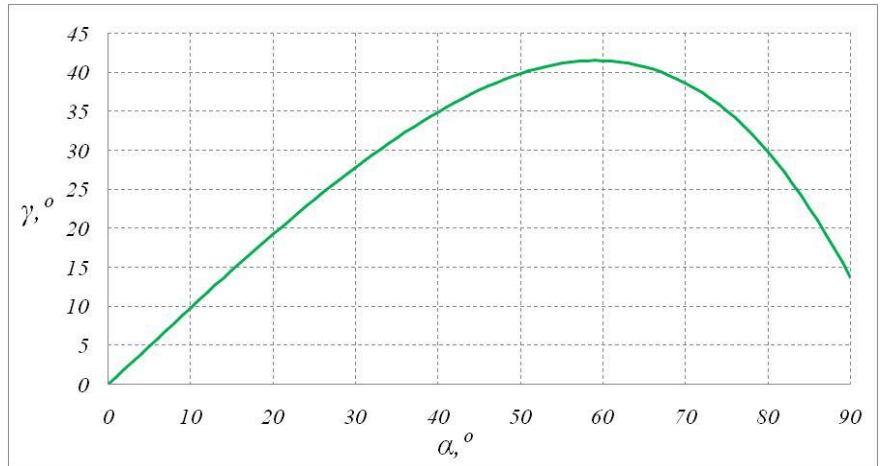

*Fig. 2. Dependence of the angle of deviation on the angle of incidence*

Since $\gamma(\alpha, n)$ is an odd function with respect to $\alpha$, it is sufficient to consider it in the region $\alpha \geq 0$. The calculations assumed that the refractive index was $n = 250/187 \approx 1.33680$, which is precisely the value used by Descartes.

It follows from the graph that all rays deflected by the drop lie in a sector of angles $\pm\gamma_{max}$. We find the angle of maximum deflection by equating the derivative of expression (1) to zero.

$$\frac{d\gamma}{d\alpha} = 2\left(2\frac{\cos\alpha}{\sqrt{n^2 - \sin^2\alpha}} - 1\right) = 0 \ .$$

It follows that the extremum is achieved if the impact angle is equal to

$$\alpha_0 = \pm \arcsin\sqrt{\frac{4-n^2}{3}} \ .$$

In this case, the maximum deviation of the light beam by a spherical drop is

$$\gamma_{01} = \pm 2\left(2\arcsin\left(\sqrt{\frac{4-n^2}{3n^2}}\right) - \arcsin\sqrt{\frac{4-n^2}{3}}\right). \tag{2}$$

Substituting $n = 250/187$, we obtain the numerical values of these parameters

$$\alpha_0 \approx 59°11', \ \delta \approx 0.859, \ \gamma_{01} \approx 41°31'.$$

Consider a thousand parallel and equidistant rays toward the upper hemisphere. Deflected by the droplet, they will cease to be parallel and unevenly fill the sector of angles $0 \le \gamma \le \gamma_1$. We divide the sector of angles *0-42°* into subsectors with a step of *2°* and count the number of rays falling into each of them. We will present the results of our calculations in the form of histograms, Fig. 3.

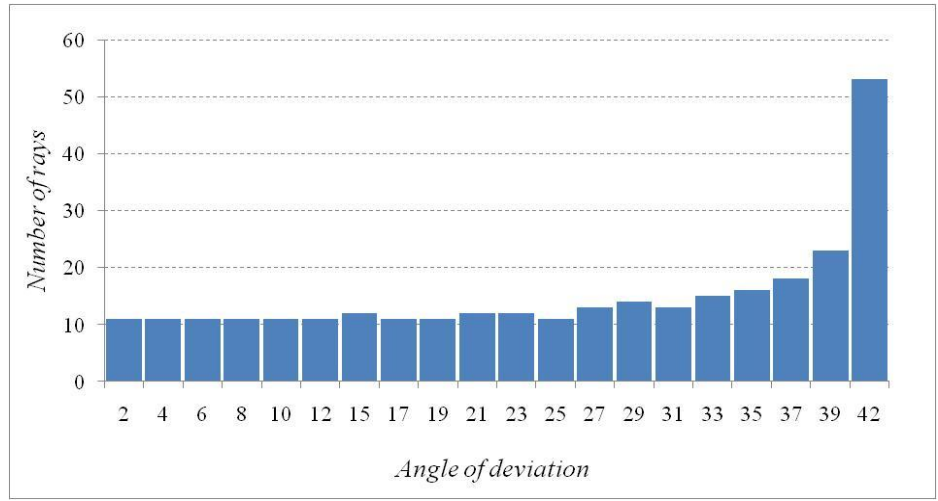

*Fig. 3. Sectoral distribution of deflected rays*

The deflected rays are concentrated near the ray that experienced the greatest deflection, called the Cartesian ray. They provide sufficient intensity for observation.

Rays falling on the lower hemisphere, upon a single reflection, will be deflected in the other direction, filling the sector of angles $-\gamma_{01} \leq \gamma \leq 0$.

### The path of a light beam in a drop with double reflection

We now follow the beam that has experienced double reflection in the droplet. Consider the beam incident on the lower hemisphere, Fig. 4. Reflected at point *C*, it, having undergone refraction at point *D*, exits the droplet in the direction opposite to the incident beam.

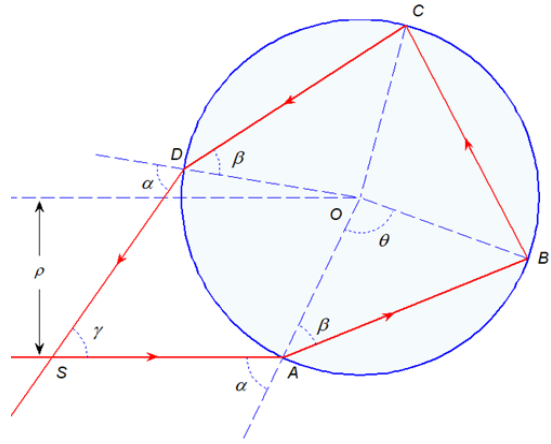

*Fig. 4. Double reflection of a beam in a spherical drop*

We calculate the angle between the direction of the source and the beam emerging from the droplet after double reflection. In the figure, it is denoted by *γ*.

Angles *AOB*, *BOC*, *COD* are equal $\theta = \pi - 2\beta$, then angle *DOA* is equal $2\pi - 3\theta = 6\beta - \pi$. Angles *ODS* and *SAO* are equal $\pi - \alpha$. For quadrilateral *AODS*, the following holds:

$$\gamma + 2(\pi - \alpha) + (6\beta - \pi) = 4\pi,$$

then

$$\gamma = 3\pi - 6\beta + 2\alpha.$$

In this case, the reflected beam is measured clockwise from the direction of the source. We'll count it counterclockwise, as in a single reflection. To do this, we'll change the sign of *α* in the resulting expression. And we'll decrease the angle by the period $2\pi$. Then

$$\gamma = \pi + 6\arcsin(\sin\alpha/n) - 2\alpha. \qquad (3)$$

Dependencies (1) and (3) for are $n = 250/187$ are plotted in Fig. 5.

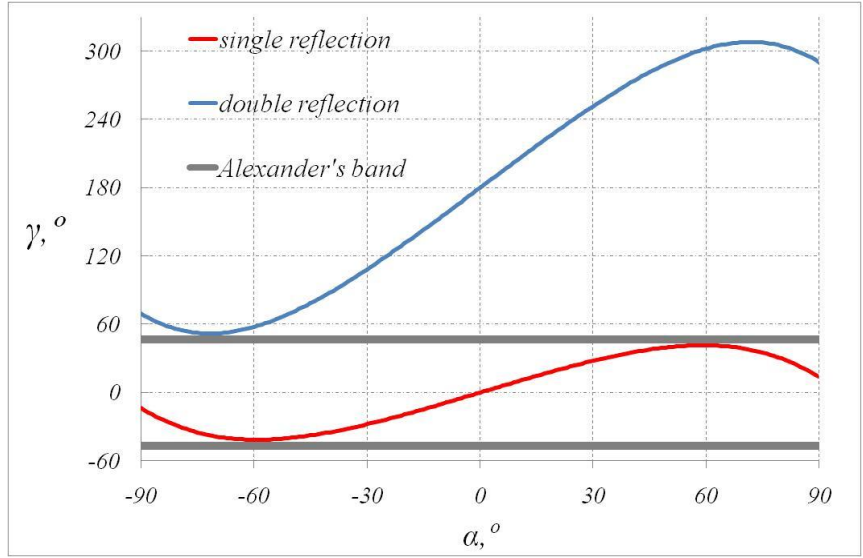

*Fig. 5. The angle between the direction to the source and the beam that has experienced one or two reflections in the droplet*

Function (3) has two extrema in its domain $-\pi/2 \leq \alpha \leq \pi/2$. Let's find them. From the condition

$$\frac{d\gamma}{d\alpha} = 2\left(3\frac{\cos\alpha}{\sqrt{n^2 - \sin^2\alpha}} - 1\right) = 0$$

it follows that the extreme is realized at

$$\alpha_0 = \pm\arcsin\sqrt{\frac{9-n^2}{8}}.$$

Then,

$$\gamma_{02} = \pi \pm \left(6\arcsin\sqrt{\frac{9-n^2}{8n^2}} - 2\arcsin\sqrt{\frac{9-n^2}{8}}\right). \qquad (4)$$

Substituting $n = 250/187$, we obtain the numerical values of these parameters

$$\alpha_0 \approx 71°43', \ \delta \approx 0.95, \ \gamma_{02} \approx 51°54'.$$

If a parallel beam of rays falls on a sphere, the rays that experience a single reflection inside the drop will be concentrated in a sector of angles $-\gamma_{01} \leq \gamma \leq \gamma_{01}$,

Fig. 5, while the rays that experience a double reflection inside the drop will be concentrated in a sector $\gamma_{02} \leq \gamma \leq 2\pi - \gamma_{02}$. The Alexander's dark band consists of sectors $\gamma_{01} \leq \gamma \leq \gamma_{02}$ and $-\gamma_{02} \leq \gamma \leq -\gamma_{01}$. Not a single scattered ray reaches it.

Let's construct a radiation pattern for radiation transmitted through a spherical drop. Assume that *401* equally spaced parallel rays are incident on the drop. We calculate the angles between the centerline and the rays reflected once and twice by the drop and distribute them into sectors with an angular size of *3°*, for a total of *120* sectors. The result is shown in Figure 6.

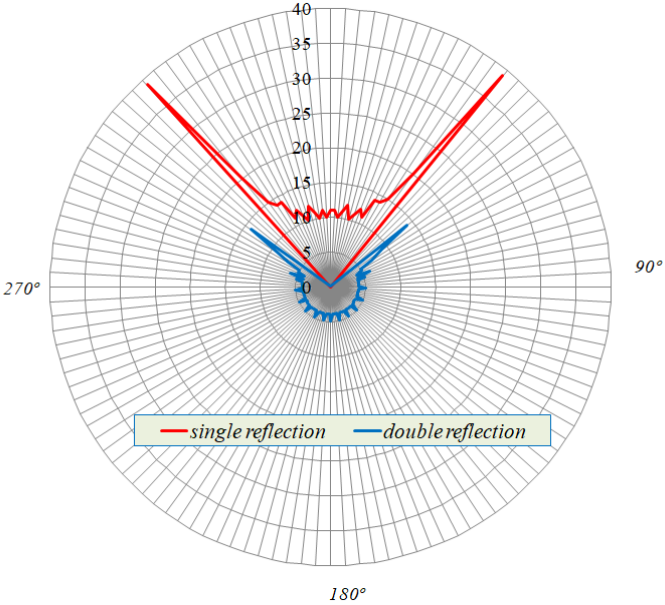

*Fig. 6. Directional pattern of radiation transmitted through a drop*

The rays are grouped in directions determined by angles $\gamma_{01}$ and $\gamma_{02}$. The region $2\pi - \gamma_{01} < \gamma < \gamma_{01}$ is brighter than the region $\gamma_{02} < \gamma$. Rays that have experienced a single reflection form the primary rainbow, while rays that have experienced a double reflection form the secondary rainbow.

With each reflection, the light intensity decreases. Therefore, the proportion of light energy that goes into rainbows of higher orders is insignificant, making them virtually undetectable.

## Newton's Theory of the Colored Rainbow

Descartes' rainbow is monochrome. While it correctly describes the redistribution of light, it does not explain how the color spectrum of this optical phenomenon is formed. The explanation for the formation of a colored rainbow, first proposed by Newton, is based on the phenomenon of dispersion – the dependence of the refractive index on wavelength.

The empirical dependence of the refractive index on the wavelength of light in transparent media, known as the Cauchy dispersion equation, has the form:

$$n = A + B/\lambda^2 + C/\lambda^4 + ... \qquad (5)$$

The dependence of the refractive index of water on the wavelength in the visible spectrum is shown in Fig. 7: *(a)* – experimental data from the reference book [8], *(b)* – the approximation curves. For distilled water in the visible spectrum, the Cauchy dispersion equation can be limited to two terms. For the given data, we obtain *A=1.32403*, *B=3080*, if the wavelength is given in nm.

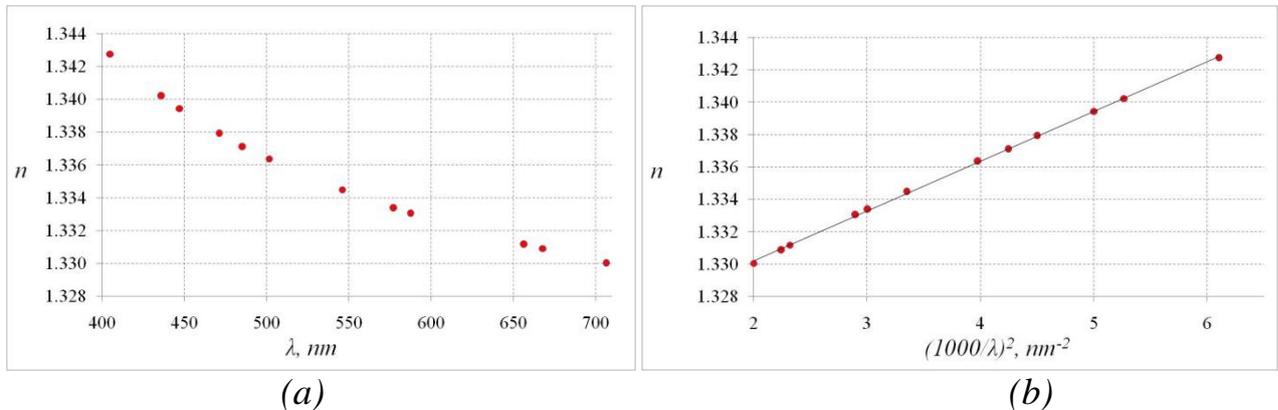

*(a)*            *(b)*
*Fig. 7. Dispersion of the refractive index of water*

Formulas (1) and (3), which determine the direction of a light beam passing through a droplet, contain the refractive index. Therefore, the directions along which light energy is concentrated differ for light waves of different wavelengths.

The value of the refractive index of water $n = 250/187$ used by Descartes corresponds to a wavelength of 490 nm, which is the blue-green region of the spectrum.

Assuming that sunlight is a mixture of monochromatic light waves – from violet, with a refractive index of $n = 109/81 \approx 1.3457$, to red, $n = 4/3 \approx 1.3333$, Newton calculated the angular characteristics of the rainbow.

A single-reflected violet beam emerging from a droplet will deviate from the direction of the light source by an angle of *40°16'*, while a red beam will deviate by *42°02'*. All other colored beams will be within this range.

When reflected twice, the violet ray will be deflected by *54°10'*, and the red ray by *50°59'*. The order of colors in the primary and secondary rainbow, Fig. 8, is inverted.

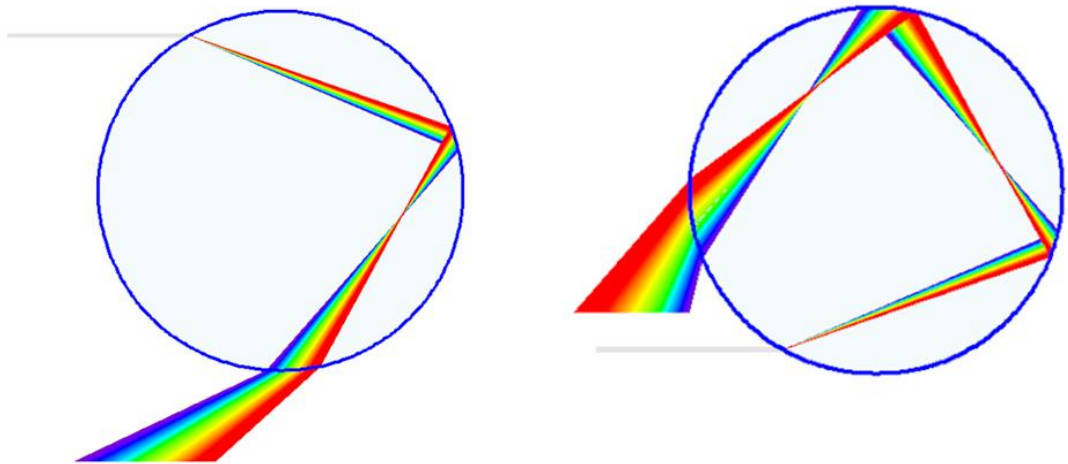

*Fig. 8. Decomposition of sunlight in a droplet*

According to (5), the refractive index $n \approx 1.3457$ corresponds to a wavelength of 377 nm, and the refractive index $n \approx 1.3333$ corresponds to a wavelength of 576 nm. According to modern concepts, violet color is a range of wavelengths of 380-440 nm, and red color – 620-740 nm. Let's assume that the red color corresponds to a wavelength of 700 nm, then, according to (5), the refractive index is 1.3303, violet – a wavelength of 400 nm, the refractive index is 1.3433. Then for the primary rainbow we obtain refined values of angles of *40°53'* and *42°28'*, and for the secondary – *50°11'* and *53°03'*. Differences from Newton's results do not exceed 2%.

Considering that a droplet in the atmosphere is struck not by a single ray, but by a cylindrical beam of white solar rays, a family of colored cones nested within one another is formed after the droplet passes through. Figure 9 shows the cones of the primary rainbow.

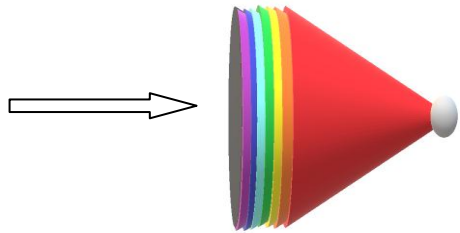

*Fig. 9. Formation of colored cones*

The colored cones formed by double reflection in a drop encompass the cones of the primary rainbow.

### Rainbow Observation Geometry

Beauty is in the eyes of the beholder. In the case of the rainbow, this expression takes on an absolute character. Let us add an observer to the scheme consisting of a light source and a drop. The correct location of the light source, the observer, and the raindrops is crucial.

Assume that the Sun is behind the observer, and the atmosphere in front of him is saturated with raindrops. Figure 10 shows only two of them.

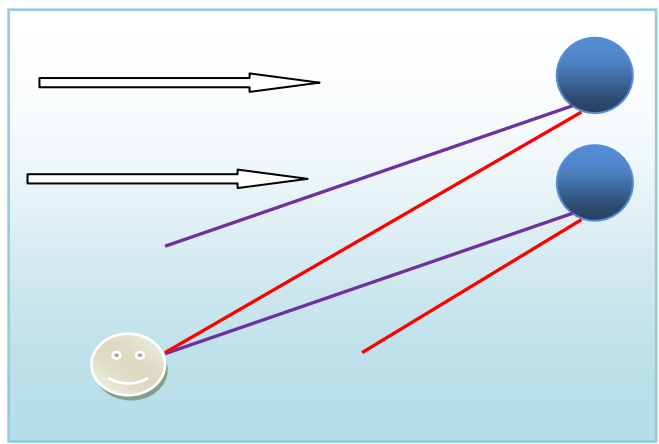

*Fig. 10. Rainbow observation scheme*

Multicolored light rays reflected once inside a droplet will be deflected by *40-42°* degrees from the direction of the sun. Only one colored ray from each droplet reaches the observer. The figure shows only red and violet rays. The drops located between the top and bottom drops will form the remaining colored rays of the rainbow spectrum. Merging, the rays from multiple drops form a rainbow arc for the observer, in which colors change from violet to red.

Similarly, rays that have experienced double reflection will be deflected at angles of *50-53°* and form a secondary band with an inverted color arrangement, Fig. 11. A dark Alexander's band is formed between the primary and secondary bands.

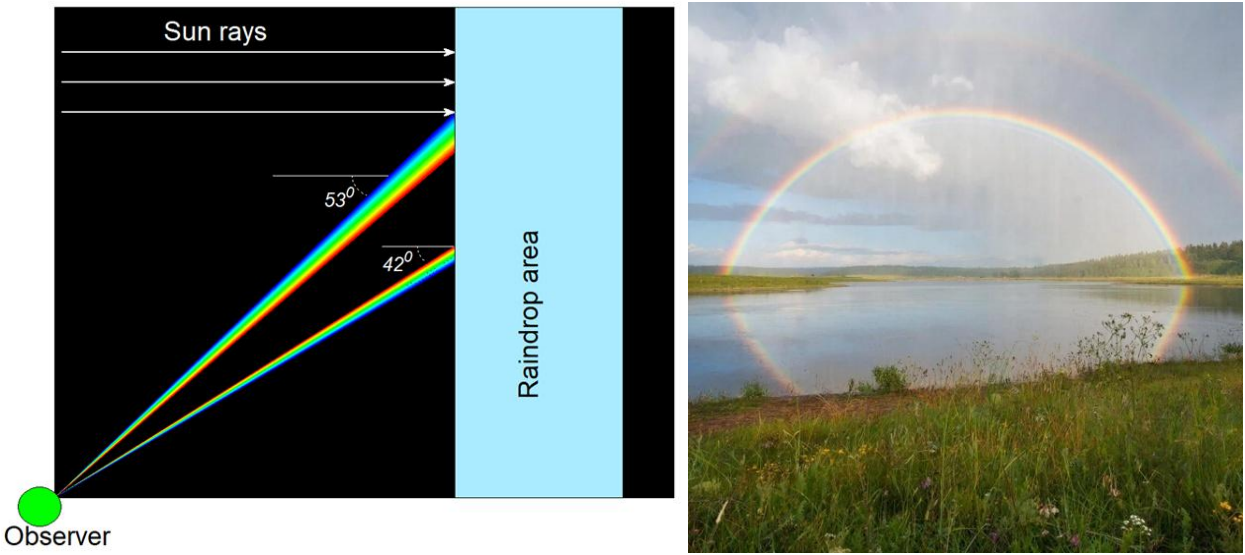

*Fig. 11. Double rainbow observation scheme*

The plane of the drawing does not necessarily have to be vertical. The path of the rays will remain the same if it is tilted from the vertical. At an angle of $\pm\pi/2$, the colored stripe will describe a semicircle, forming a rainbow arc, Fig. 11.

If the Sun rises above the horizon by more than *53°03′*, the maximum deflection of the violet ray during double reflection, this ray will become horizontal, meaning it and all other rays will not reach the observer's eye. If the observer ascends in an aircraft, they will be able to see a rainbow. And this rainbow will not be an arc, but a full circle.

In reality, the sun's rays are not strictly parallel. The angular size of the Sun for an observer on Earth is ~32 arcmin. Therefore, the light rays arriving from the droplet to the observer are not strictly monochromatic. This reduces the rainbow's contrast, but the fundamental pattern remains.

## Rainbow on Venus

The dense clouds of Venus consist primarily of sulfur dioxide and droplets of sulfuric acid [9]. The mechanism for the formation of rainbows under these conditions is similar to that on Earth. Only the dispersion dependence of the refractive index changes.

The results of measurements of the optical properties of sulfuric acid solutions with a mass concentration of 95.6, 84.5, 75.0, 50.0, 38.0 and 25.0% are given in [10].

Table 1 contains the refractive index values in the visible and near ultraviolet ranges at different acid concentrations.

Table 1

| *nm* | *0.0%* | *25.0%* | *38.0%* | *50.0%* | *75.0%* | *84.5%* | *95.6%* |
|---|---|---|---|---|---|---|---|
| *360* | *1.3518* | *1.3830* | *1.4070* | *1.4210* | *1.4520* | *1.4630* | *1.4590* |
| *408* | *1.3422* | *1.3730* | *1.3920* | *1.4080* | *1.4380* | *1.4480* | *1.4430* |
| *449* | *1.3387* | *1.3690* | *1.3870* | *1.4020* | *1.4320* | *1.4420* | *1.4380* |
| *556* | *1.3341* | *1.3660* | *1.3840* | *1.3970* | *1.4310* | *1.4380* | *1.4340* |
| *702* | *1.3295* | *1.3630* | *1.3820* | *1.3940* | *1.4280* | *1.4360* | *1.4320* |
| *714* | *1.3292* | *1.3630* | *1.3810* | *1.3940* | *1.4270* | *1.4350* | *1.4320* |

Let us present the measurement results in the form of graphs, Fig. 12. In the visible part of the spectrum, the behavior of the dispersion curve for different concentrations of sulfuric acid does not change fundamentally.

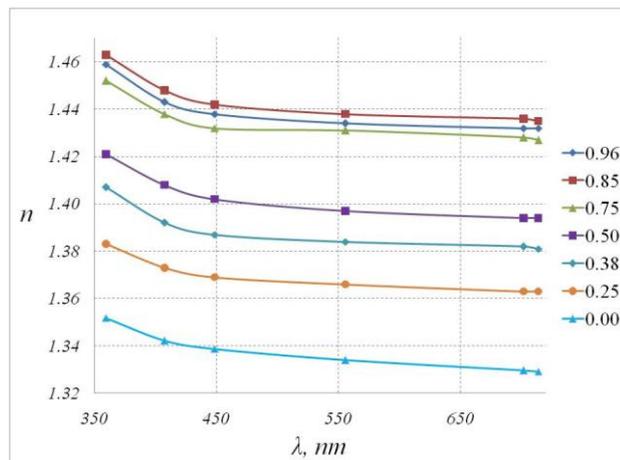

*Fig. 12. Dispersion of visible light in sulfuric acid solutions*

Figure 13 shows the results of refractive index measurements for red (702 nm) and violet (408 nm) light at different acid mass fractions.

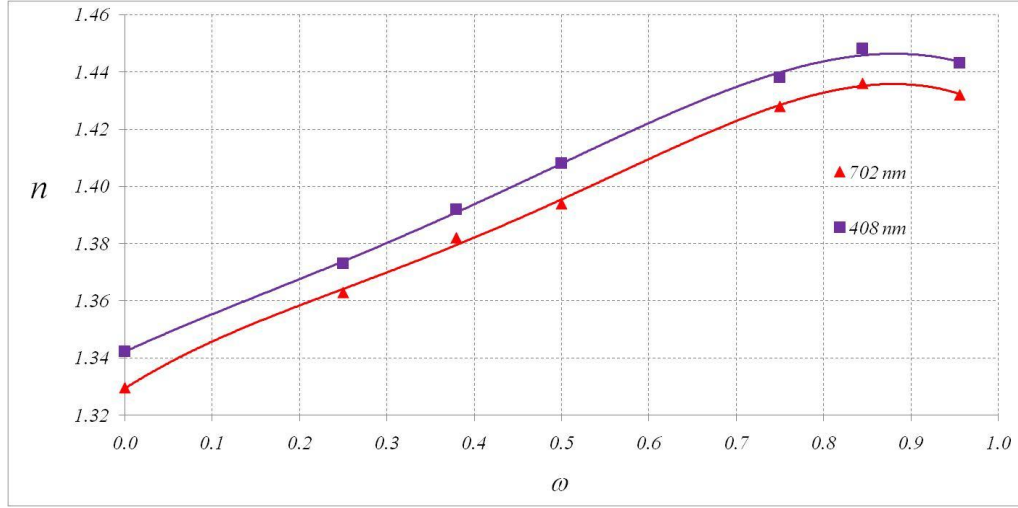

*Fig. 13. Refractive index versus acid mass fraction*

Let $\omega$ be the acid mass fraction. Approximating this with a fourth-order polynomial yields:

$$n = C_0 + C_1\omega + C_2\omega^2 + C_3\omega^3 + C_4\omega^4.$$

The polynomial coefficients are listed in Table 2.

Table 2

| nm | $C_0$ | $C_1$ | $C_2$ | $C_3$ | $C_4$ |
|---|---|---|---|---|---|
| 702 | 1.329420 | 0.194585 | -0.374914 | 0.722222 | -0.444778 |
| 408 | 1.342120 | 0.140586 | -0.128891 | 0.359853 | -0.274816 |

Let's calculate the angles of maximum intensity for the primary $\gamma_{01}$ and secondary $\gamma_{02}$ rainbows for the red and violet rays as a function of the mass fraction of sulfuric acid. The calculation results are shown in the figure (Fig. 14).

With an increase in the mass fraction of sulfuric acid from *0* to *90%* for red, the angle $\gamma_{01}$ decreases by *0.7* times from *42°36'* to *29°13'*, the angle $\gamma_{02}$ increases by *1.5* times from *49°57'* to *74°35'*. The angular size of the Alexander's band increases by more than six times, from *7°20'* to *45°23'*.

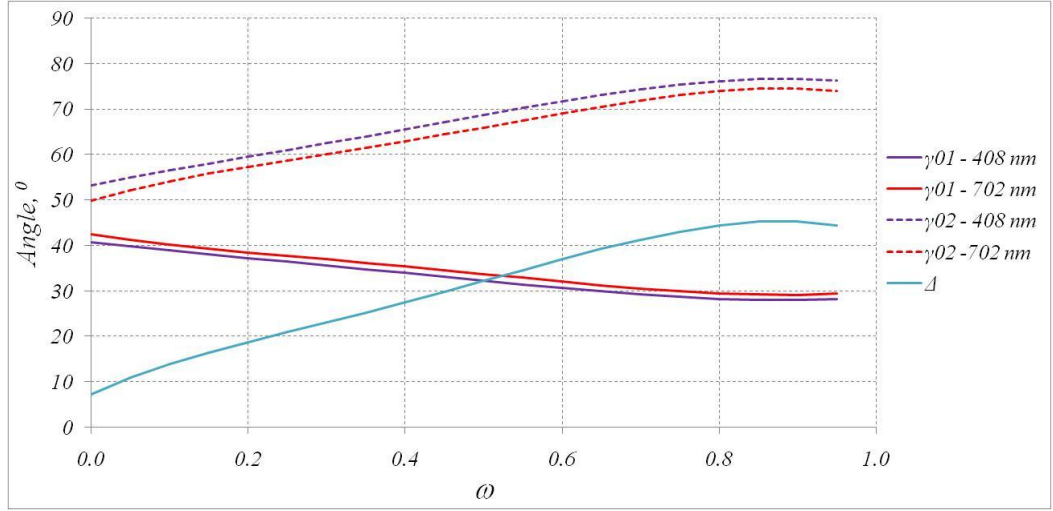

*Fig. 14. Dependence of the refractive index on the acid mass fraction*

If we measure the angular size of the Alexander's band in degrees, then by approximating the calculation results with a polynomial, we obtain

$$\Delta = D_0 + D_1\omega + D_2\omega^2 + D_3\omega^3 + D_4\omega^4,$$

where $D_0=7.4253$, $D_1=75.8766$, $D_2=-137.4143$, $D_3=242.6528$, $D_4=-146.1655$.

## Conclusion

The formation mechanism of rainbows is examined within the framework of geometric optics. Quantitative estimates of the influence of sulfuric acid concentration in raindrops on rainbow parameters are provided. Analysis of the angular distribution of light intensity in the rainbow allows one to determine the cloud's composition. The angular size of the Alexander's band is particularly sensitive to changes in sulfuric acid concentration. This effect may be used as a remote diagnostic of sulfuric acid concentration in Venus clouds.

# РАДУГА ОТ ДЕКАРТА ДО ВЕНЕРЫ

А.Д. Заикин

Новосибирский государственный технический университет, г. Новосибирск, Россия

Рассмотрен механизм образования радуги по Декарту и Ньютону. В рамках геометрической оптики рассчитаны параметры дождевой радуги на Венере. В предположении, что солнечные лучи преломляются на сферических каплях, состоящих из водного раствора серной кислоты, получены зависимости угловых размеров первичной и вторичной радуги от концентрации раствора. Показано, что наиболее сильно изменение концентрации сказывается на размерах полосы Александра, угловом расстоянии между первичной и вторичной радугой.

**Ключевые слова:** радуга, Декарт, геометрическая оптика, дисперсия света, серная кислота, Венера.

## Введение

Радуга, наблюдаемая, когда Солнце освещает завесу дождя, – весьма распространенное оптическое явление. Наряду с эстетическим наслаждением она порождает у наблюдателя закономерные вопросы о причинах ее появления. Первая научная публикация о механизме образования радуги, принадлежащая Декарту, увидела свет в 1637 году. Применив законы геометрической оптики, он объяснил [1] возникновение радуги преломлением солнечных лучей в каплях воды. Основываясь на явлении дисперсии света, Ньютон прояснил [2] механизм образования цветной радуги. Волновая теория света работами Юнга и Эйри позволила описать тонкие эффекты образования в радуге дуг высших порядков. Точное решение, построенное Ми, основано на расчете рассеяния электромагнитной волны на сферическом включении. Обзорные работы [3,4] дают представление о волновой теории радуги.

Наличие точного численного решения не сделало тему радуги полностью исчерпанной. Зачастую анализ таких решений оказывается сложнее, чем его построение. Остается актуальным анализ физической сущности наблюдаемых в атмосфере разнообразных оптических явлений, родственных классической дождевой радуге.

Очевидцы, наблюдавшие полет Тунгусского космического тела, отмечали формирование радужного следа вдоль его траектории. С точки зрения авторов, изложенной в [5], наличие радужного следа является свидетельством того, что данное космическое тело могло быть только кометой, из которой непосредственно во время полета в мезопаузе выделялась вода.

Глория – цветные кольца света на облаках вокруг тени наблюдателя, находящегося на горе или самолете. Источник света (Солнце или Луна) при этом располагается за наблюдателем. Если дождевая радуга возникает из-за преломления света, то глория – результат дифракции света на очень мелких и однородных каплях.

В [6] показано, что глория формируется в облаках с отрицательными температурами на жидких сферических частицах аморфной воды с показателем преломления 1.81-1.82 и диаметром более 20 мкм. Геометрические характеристики глории позволяют дистанционно получать информацию о дисперсных фазах в холодных облаках.

Отдельно стоит отметить задачи о радужных явлениях во внеземных атмосферах. 24 июля 2011 года орбитальным аппаратом VenusExpress ЕКА на высоте 70 км над поверхностью Венеры было зафиксировано [7] радужное явление (глория). Для наблюдения глории требовалось специальное позиционирование исследовательского аппарата, он должен находиться прямо между Солнцем и облаком, отражающим солнечный свет. Глория находит свое объяснение в рамках дифракции света.

В настоящей работе ограничимся приближением геометрической оптики, которое корректно описывает угловые положения экстремумов радуги. Проведем анализ параметров классической радуги, возникающей при рассеянии света на каплях серной кислоты.

### Ход светового луча в капле при однократном отражении

На сферическую каплю радиуса $R$ падает луч света. Расстояние от луча до параллельной ему прямой, проходящей через центр сферы, называется прицельным. Обозначив прицельное расстояние как $\rho$, определим также безразмерный прицельный параметр $\delta = \rho/R$. Угол между лучом и нормалью к поверхности сферы в точке падения назовем углом падения и обозначим как *α*. Полагаем, что источник параллельных лучей (Солнце), луч, центр капли и глаз наблюдателя находятся в одной вертикальной плоскости. Наблюдатель, находясь между Солнцем и каплей, располагается ниже ее.

Рассмотрим луч, падающий на верхнюю полусферу (Рис. 1). Преломившись в точке *A* под углом *β*, луч достигнет поверхности сферы в точке *B*, отразится и, преломившись в точке *C*, выйдет из сферы в направлении, противоположном падающему лучу. В точках *A* и *C* произойдет также отражение луча, а в точке *B* – преломление, однако в данной постановке эти явления несущественны.

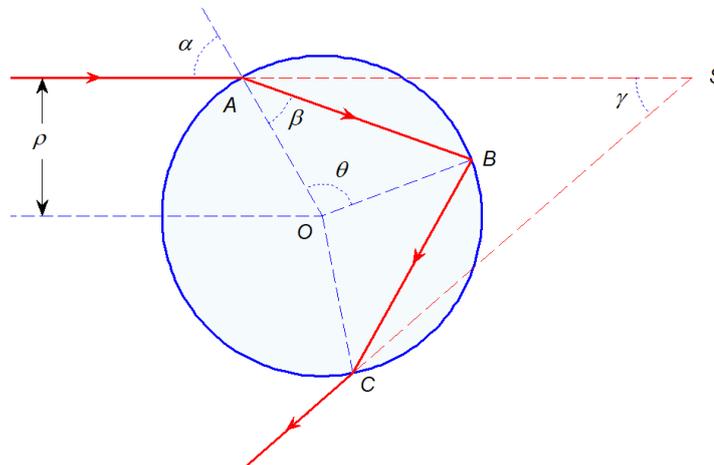

*Рис. 1. Однократное отражение луча в сферической капле*

Так как световой луч может падать и на нижнюю полусферу, то значения угла *α* лежат в диапазоне $\pm\pi/2$, а $-1 \leq \delta \leq 1$, поскольку $\delta = \sin\alpha$.

Угол падения и угол отражения в точке *B* равны. Треугольники *AOB* и *BOC* равнобедренные, поэтому все четыре угла при их основаниях равны *β*. Угол, образованный радиусами в этих треугольниках, обозначим как *θ*.

Продолжим падающий на сферу и вышедший из нее лучи до пересечения в точке *S*. Вышедший из капли луч отклонится от направления на источник света на угол *γ*. Будем отсчитывать его против часовой стрелки от направления на источник.

Для равнобедренного треугольника *ASC* справедливо

$$AC^2 = AS^2 + SC^2 - 2AS \cdot SC \cdot \cos\gamma .$$

Поскольку $AC = 2R\sin\theta$, то

$$AS = \frac{AC}{\sqrt{2(1-\cos\gamma)}} = \frac{\sqrt{2}R\sin\theta}{\sqrt{1-\cos\gamma}} .$$

Для треугольника $AOB$ справедливо $\theta + \beta + \beta = \pi$, а для треугольника $AOS$ $\theta + \alpha + \gamma/2 = \pi$. Вычитая эти соотношения, получаем, что

$$\gamma = 2(2\beta - \alpha).$$

Пусть $n$ – абсолютный показатель преломления материала капли (в нашем случае воды). Абсолютный показатель преломления воздуха положим равным единице, тогда из закона Снеллиуса следует, что

$$\sin\alpha = n\sin\beta, \quad \beta = \arcsin(\sin\alpha/n) ,$$

а угол отклонения луча равен

$$\gamma = 4\arcsin(\sin\alpha/n) - 2\alpha . \qquad (1)$$

График построенной функции $\gamma(\alpha, n)$ приведен на рисунке (Рис. 2).

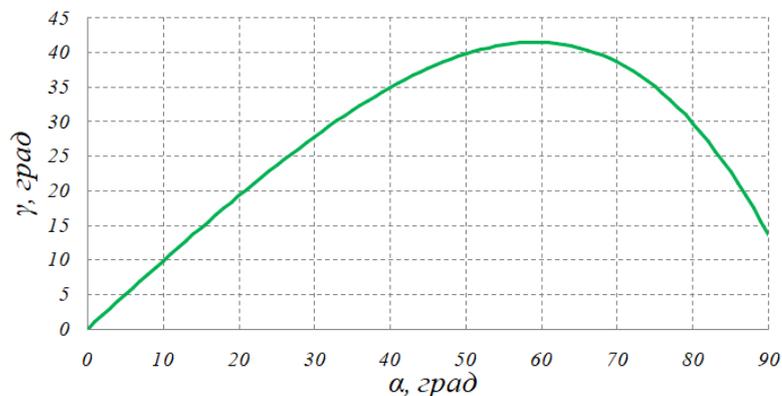

*Рис. 2. Зависимость угла отклонения от угла падения*

Поскольку $\gamma(\alpha, n)$ – нечетная функция относительно *α*, достаточно рассмотреть ее в области $\alpha \geq 0$. В расчетах полагалось, что абсолютный показатель преломления $n = 250/187 \approx 1.33680$, именно такое значение использовал Декарт.

Из графика следует, что все отклоненные каплей лучи лежат в секторе углов $\pm\gamma_{\max}$. Найдем угол максимального отклонения, приравняв производную от выражения (1) нулю

$$\frac{d\gamma}{d\alpha} = 2\left(2\frac{\cos\alpha}{\sqrt{n^2 - \sin^2\alpha}} - 1\right) = 0 \quad .$$

Отсюда следует, что экстремум достигается, если прицельный угол равен

$$\alpha_0 = \pm\arcsin\sqrt{\frac{4-n^2}{3}} \quad .$$

При этом максимальное отклонение светового луча сферической каплей составляет

$$\gamma_{01} = \pm 2\left(2\arcsin\left(\sqrt{\frac{4-n^2}{3n^2}}\right) - \arcsin\sqrt{\frac{4-n^2}{3}}\right). \tag{2}$$

Подставив $n = 250/187$, получим численные значения этих параметров

$$\alpha_0 \approx 59°11', \; \delta \approx 0.859, \; \gamma_{01} \approx 41°31'.$$

Направим на верхнюю полусферу тысячу параллельных и равноудаленных лучей. Они, отклоненные каплей, перестанут быть параллельными, и неравномерно заполнят сектор углов $0 \le \gamma \le \gamma_1$. Разобьем сектор углов *0-42°* на подсекторы с шагом *2°* и подсчитаем число лучей, попадающих в каждый из них. Результаты подсчетов представим в виде гистограммы (Рис. 3).

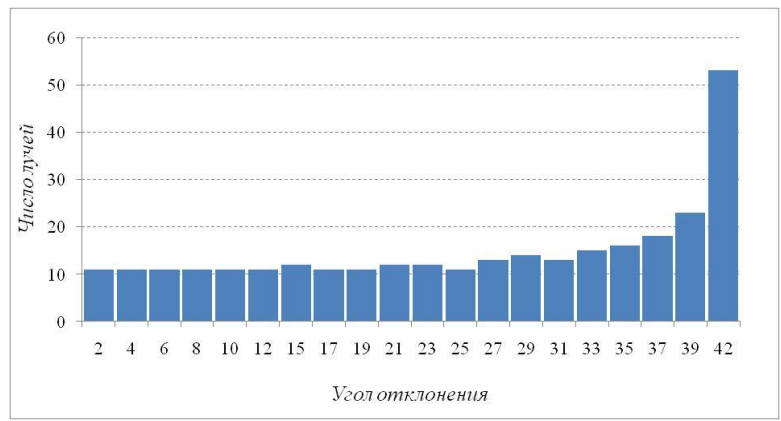

*Рис. 3. Секторальное распределение отклоненных лучей*

Отклоненные лучи концентрируются вблизи луча, испытавшего максимальное отклонение, называемым лучом Декарта. Именно они обеспечивают суммарную интенсивность, достаточную для наблюдения.

Лучи, падающие на нижнюю полусферу, при однократном отражении отклонятся в другую сторону, заполняя сектор углов $-\gamma_{01} \leq \gamma \leq 0$.

**Ход светового луча в капле при двукратном отражении**

Теперь проследим за лучом, испытавшим двукратное отражение в капле. Рассмотрим луч, падающий на нижнюю полусферу (Рис. 4). Отраженный в точке *C* он, испытав преломление в точке *D*, выйдет из капли в направлении, противоположном падающему лучу.

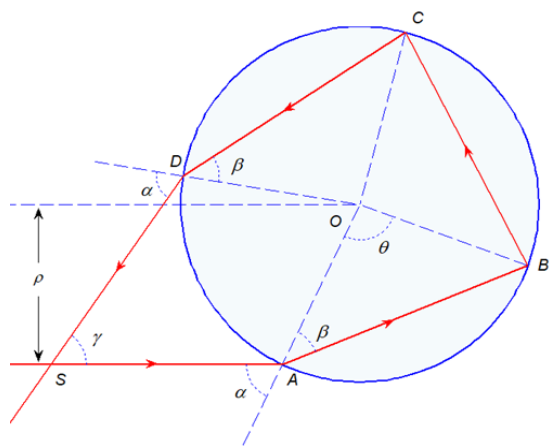

*Рис. 4. Двукратное отражение луча в сферической капле*

Рассчитаем угол между направлением на источник и вышедшим после двукратного отражения из капли лучом. На рисунке он обозначен как *γ*.

Углы *AOB*, *BOC*, *COD* равны $\theta = \pi - 2\beta$, тогда угол *DOA* равен $2\pi - 3\theta = 6\beta - \pi$. Углы *ODS* и *SAO* равны $\pi - \alpha$. Для четырехугольника *AODS* справедливо

$$\gamma + 2(\pi - \alpha) + (6\beta - \pi) = 4\pi,$$

тогда

$$\gamma = 3\pi - 6\beta + 2\alpha.$$

Отраженный луч в этом случае отсчитывается от направления на источник по часовой стрелке. Будем отсчитывать его против часовой стрелки как при однократном отражении. Для этого изменим в полученном выражении знак $\alpha$. И уменьшим угол на период $2\pi$. Тогда

$$\gamma = \pi + 6\arcsin(\sin\alpha/n) - 2\alpha. \tag{3}$$

Зависимости (1) и (3) при $n = 250/187$ представлены на рисунке (Рис. 5).

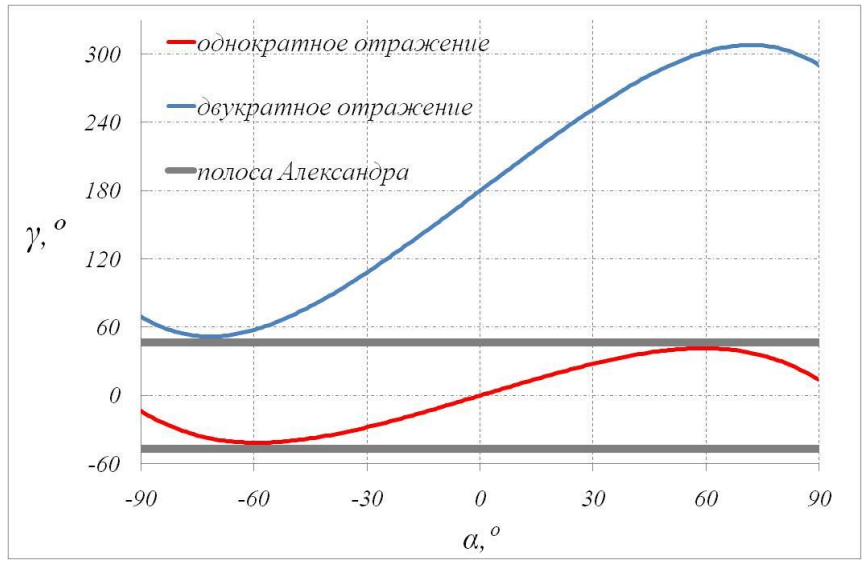

*Рис. 5. Угол между направлением на источник и лучом, испытавшим одно или двукратное отражение в капле*

Функция (3) в области определения $-\pi/2 \leq \alpha \leq \pi/2$ имеет два экстремума. Найдем их. Из условия

$$\frac{d\gamma}{d\alpha} = 2\left(3\frac{\cos\alpha}{\sqrt{n^2 - \sin^2\alpha}} - 1\right) = 0$$

следует, что экстремум реализуется при

$$\alpha_0 = \pm\arcsin\sqrt{\frac{9 - n^2}{8}}.$$

Тогда,

$$\gamma_{02} = \pi \pm \left(6\arcsin\sqrt{\frac{9 - n^2}{8n^2}} - 2\arcsin\sqrt{\frac{9 - n^2}{8}}\right). \tag{4}$$

Подставив $n = 250/187$, получим численные значения этих параметров

$$\alpha_0 \approx 71°43', \ \delta \approx 0.95, \ \gamma_{02} \approx 51°54'.$$

Если на сферу падает параллельный пучок лучей, то лучи, испытавшие однократное отражение внутри капли, окажутся сосредоточенными в секторе углов $-\gamma_{01} \leq \gamma \leq \gamma_{01}$, Рис. 5, а лучи, испытавшие двукратное отражение внутри капли, окажутся сосредоточенными в секторе $\gamma_{02} \leq \gamma \leq 2\pi - \gamma_{02}$. Полоса Александра – это сектора $\gamma_{01} \leq \gamma \leq \gamma_{02}$ и $-\gamma_{02} \leq \gamma \leq -\gamma_{01}$. Туда не попадает ни один рассеянный луч.

Построим диаграмму направленности излучения, прошедшего через сферическую каплю. Пусть на каплю падают *201* равноотстоящих параллельных лучей. Рассчитаем углы между осевой линией и лучами, одно и двукратно отраженными в капле, и распределим их по секторам с угловым размером *3°*, общим количеством *120* секторов. Результат представлен на рисунке (Рис. 6).

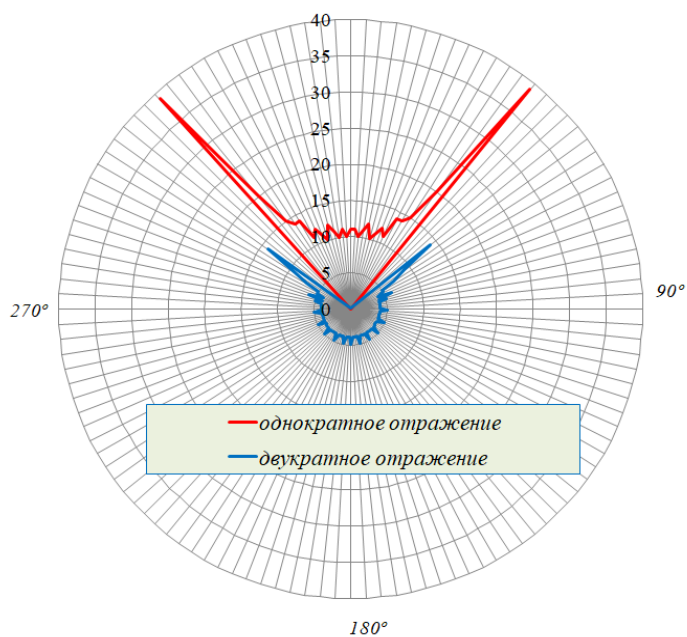

*Рис. 6. Диаграмма направленности излучения, рассеянного каплей*

Лучи группируются в направлениях, определяемых углами $\gamma_{01}$ и $\gamma_{02}$. Область $2\pi - \gamma_{01} < \gamma < \gamma_{01}$ более светлая, чем область $\gamma_{02} < \gamma$. Лучи, испытавшие однократное отражение, формируют первичную радугу, а лучи, испытавшие двукратное отражение, – вторичную.

В каждом акте отражения интенсивность света уменьшается. Поэтому доля световой энергии, приходящаяся на радуги более высоких порядков незначительна, что делает их практически ненаблюдаемыми.

**Цветная радуга Ньютона**

Радуга Декарта монохромна, правильно описывая перераспределение светового потока, она не объясняет, как формируется цветовая гамма этого оптического явления. Объяснение механизма образования цветной радуги, впервые предложенное Ньютоном, базируется на явлении дисперсии – зависимости абсолютного показателя преломления от длины волны.

Эмпирическая зависимость показателя преломления от длины волны света в прозрачных средах известная как дисперсионное уравнение Коши имеет вид:

$$n = A + B/\lambda^2 + C/\lambda^4 + ... \tag{5}$$

Зависимость абсолютного показателя преломления воды от длины волны в видимой области спектра приведена на Рис. 7: *(а)* – экспериментальные данные из справочника [8], *(б)* – построенная аппроксимирующая кривая. Для дистиллированной воды в видимой области спектра в дисперсионном уравнении Коши можно ограничиться двумя слагаемыми. Для приведенных данных получаем, что *A=1.32403, B=3080,* если длина волны задана в нм.

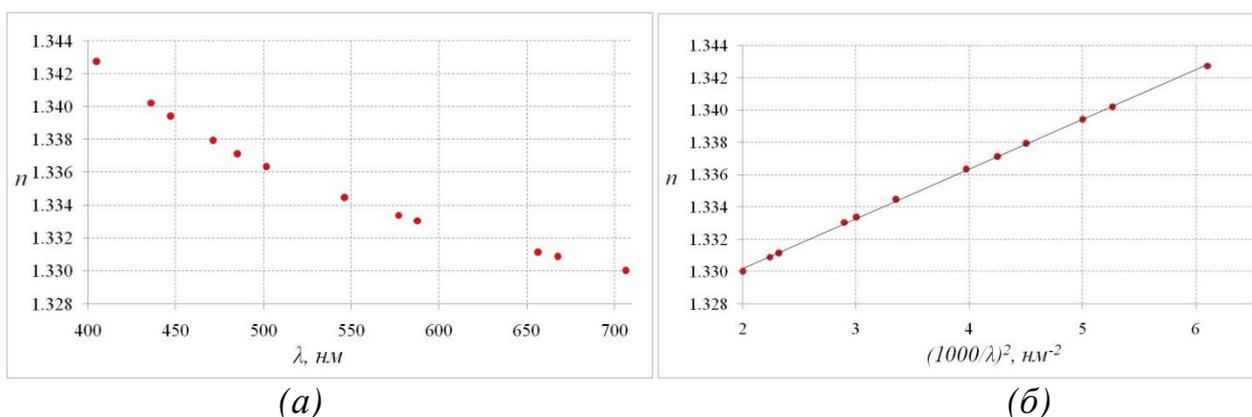

*(а)*          *(б)*

*Рис. 7. Дисперсия показателя преломления воды*

Формулы (1) и (3), определяющие направление светового луча, прошедшего через каплю, содержат абсолютный показатель преломления. Поэтому направления, вдоль которых концентрируется световая энергия, отличаются для световых волн различной длины.

Использованное Декартом значение абсолютного показателя преломления воды $n = 250/187$ соответствует длине волны 490 нм, это сине-зеленая область спектра.

Полагая солнечный свет смесью монохроматических световых волн – от фиолетового, абсолютный показатель преломления $n = 109/81 \approx 1.3457$, до красного, $n = 4/3 \approx 1.3333$, Ньютон рассчитал угловые характеристики радуги.

Испытавший однократное отражение и вышедший из капли фиолетовый луч отклонится от направления на источник света на угол, равный *40°16′*, а красный – на *42°02′*. Все остальные цветные лучи будут находиться в этом диапазоне.

При двукратном отражении фиолетовый луч отклонится на *54°10′*, а красный – *50°59′*. Порядок цветов в первичной и вторичной радуге, Рис. 8, инверсный.

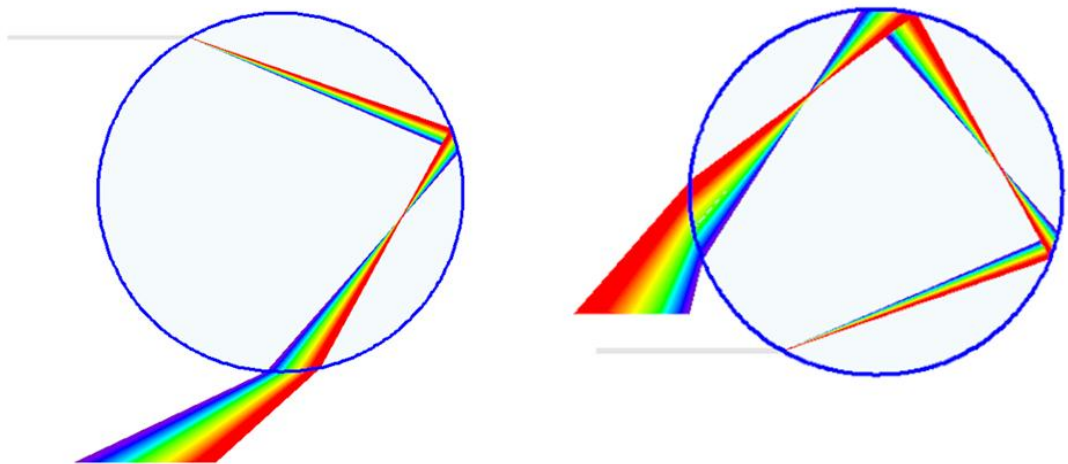

*Рис. 8. Разложение солнечного света в капле*

Согласно (5), показателю преломления $n \approx 1.3457$ соответствует длина волны 377 нм, а показателю преломления $n \approx 1.3333$ – длина волны 576 нм. По современным представлениям фиолетовый цвет – это диапазон длин волн 380-440 нм, а красный цвет – 620-740 нм. Положим, что красному цвету отвечает длина волны 700 нм, тогда, согласно (5) показатель преломления 1.3303, фиолетовому – длина волны 400 нм, показатель преломления 1.3433. Тогда для первичной радуги получаем уточненные значения углов *40°53′* и *42°28′*, а для вторичной – *50°11′* и *53°03′*. Отличия от результатов Ньютона не превышают 2%.

Если учесть, что на каплю в атмосфере падает не один луч, а цилиндрический пучок белых солнечных лучей, то после прохождения капли образуется семейство цветных конусов, вставленных друг в друга. Конусы первичной радуги показаны на рисунке (Рис. 9).

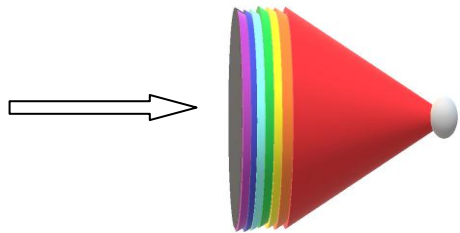

*Рис. 9. Образование цветных конусов*

Цветные конусы, образующиеся при двукратном отражении в капле, охватывают конусы первичной радуги.

### Схема наблюдения радуги

Красота в глазах смотрящего. В случае с радугой это выражение приобретает абсолютный характер. Дополним схему, состоящую из источника света и капли, наблюдателем. Правильное расположение источника света, наблюдателя и дождевых капель имеет принципиальное значение.

Пусть Солнце находится за спиной наблюдателя, а атмосфера перед ним насыщена дождевыми каплями. На Рис. 10 изображены лишь две из них

Испытавшие однократное отражение внутри капли разноцветные световые лучи отклонятся на *40-42°* от направления на Солнце. При этом от каждой капли до наблюдателя доходит только один цветной луч. На рисунке представлены только красный и фиолетовый лучи. Капли, находящиеся между верхней и нижней каплей, сформируют остальные цветные лучи радужного спектра. Сливаясь, лучи от множества капель образуют для наблюдателя радужную полоску, в которой цвета меняются от фиолетового до красного.

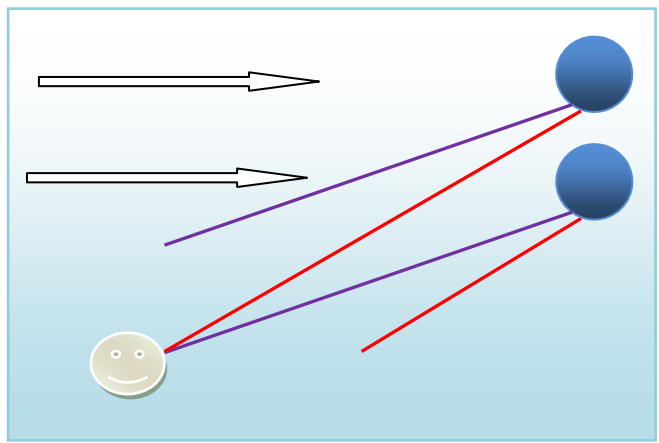

*Рис. 10. Схема наблюдения радуги*

Аналогично лучи, испытавшие двукратное отражение, отклонятся на углы *50-53°* и образуют вторичную полоску с инверсным расположением цветов, Рис. *11*. Между первичной и вторичной полосами образуется темная полоса Александра.

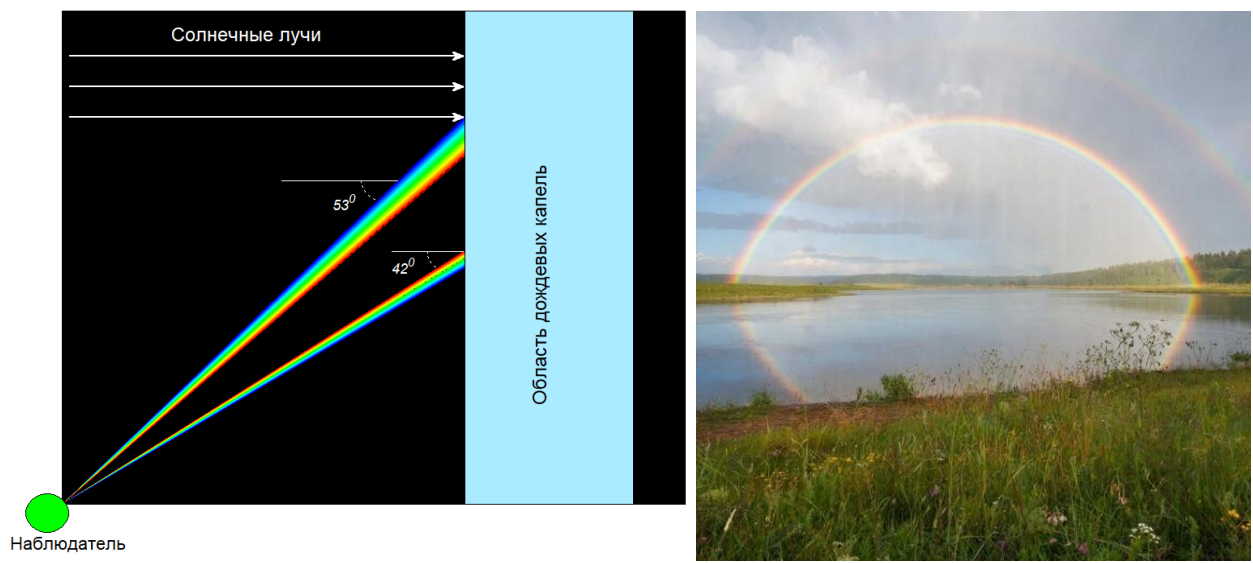

*Рис. 11. Схема наблюдения двойной радуги*

Плоскость рисунка не обязательно должна быть вертикальной. Ход лучей останется прежним, если отклонить ее от вертикали. При отклонении на угол $\pm\pi/2$ цветная полоска опишет полуокружность, сформировав дугу радуги, Рис. 11.

Если Солнце поднимется над горизонтом более чем на *53°03'*, а это максимальное отклонение фиолетового луча при двукратном отражении, то этот луч станет горизонтальным, а значит, он и все остальные лучи не попадут в глаз наблюдателя. Если же наблюдатель поднимется на летательном аппарате, то он сможет увидеть радугу. И радуга эта будет не дугой, а полной окружностью.

В действительности солнечные лучи не строго параллельны. Угловой размер Солнца для земного наблюдателя составляет ~32′. Поэтому к наблюдателю от капли приходит не строго монохроматические лучи света. Это уменьшает контрастность радуги, но принципиально картина сохраняется.

### Радуга на Венере

Плотные облака Венеры состоят в основном из диоксида серы и капель серной кислоты [9]. Механизм формирования дождевой радуги в таких условиях аналогичен земному. Изменяется лишь дисперсионная зависимость абсолютного показателя преломления.

Результаты измерений оптических свойств растворов серной кислоты, с концентрацией по массе 95.6, 84.5, 75.0, 50.0, 38.0 и 25.0%, приведены в [10].

Таблица 1 содержит значения абсолютного показателя преломления в видимом диапазоне и ближнем ультрафиолете при различных концентрациях кислоты.

Таблица 1

| *нм* | *0.0%* | *25.0%* | *38.0%* | *50.0%* | *75.0%* | *84.5%* | *95.6%* |
|---|---|---|---|---|---|---|---|
| *360* | *1.3518* | *1.3830* | *1.4070* | *1.4210* | *1.4520* | *1.4630* | *1.4590* |
| *408* | *1.3422* | *1.3730* | *1.3920* | *1.4080* | *1.4380* | *1.4480* | *1.4430* |
| *449* | *1.3387* | *1.3690* | *1.3870* | *1.4020* | *1.4320* | *1.4420* | *1.4380* |
| *556* | *1.3341* | *1.3660* | *1.3840* | *1.3970* | *1.4310* | *1.4380* | *1.4340* |
| *702* | *1.3295* | *1.3630* | *1.3820* | *1.3940* | *1.4280* | *1.4360* | *1.4320* |
| *714* | *1.3292* | *1.3630* | *1.3810* | *1.3940* | *1.4270* | *1.4350* | *1.4320* |

Представим результаты измерений в виде графиков, Рис. 12. В области видимой части спектра поведение кривой дисперсии для различных концентраций серной кислоты принципиально не меняется.

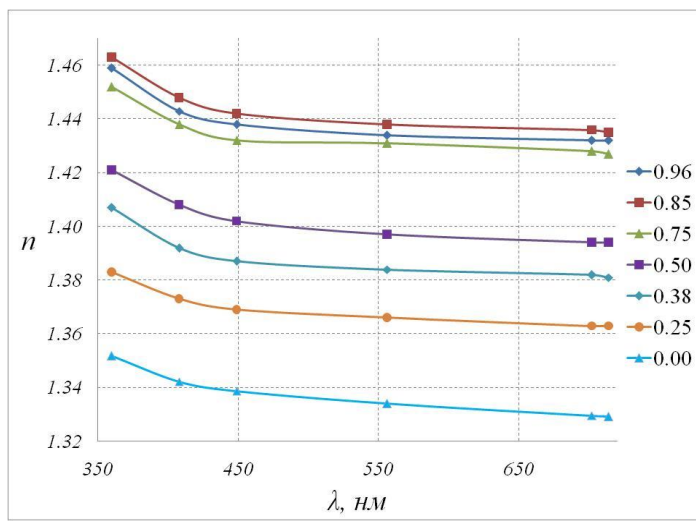

*Рис. 12. Дисперсия видимого света в растворах серной кислоты*

На рисунке (Рис. 13) представлены результаты измерений абсолютного показателя преломления для красного (702 нм) и фиолетового (408 нм) света при различных значениях массовой доли кислоты.

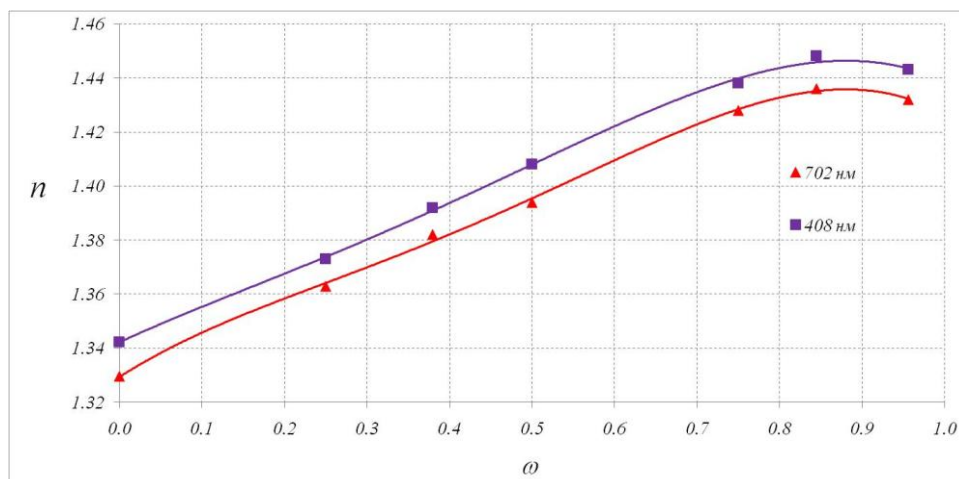

*Рис. 13. Зависимость показателя преломления от массовой доли кислоты*

Пусть $\omega$ – массовая доля кислоты, тогда аппроксимируя $n(\omega)$ полиномом четвертого порядка, получаем:

$$n = C_0 + C_1\omega + C_2\omega^2 + C_3\omega^3 + C_4\omega^4.$$

Коэффициенты полинома приведены в таблице 2.

Таблица 2

| нм | $C_0$ | $C_1$ | $C_2$ | $C_3$ | $C_4$ |
|---|---|---|---|---|---|
| 702 | 1.329420 | 0.194585 | -0.374914 | 0.722222 | -0.444778 |
| 408 | 1.342120 | 0.140586 | -0.128891 | 0.359853 | -0.274816 |

Рассчитаем углы максимальной интенсивности первичной и вторичной радуги $\gamma_{01}$ и $\gamma_{02}$ для красного и фиолетового луча как функцию массовой доли серной кислоты. Результаты расчетов приведены на рисунке (Рис. 14).

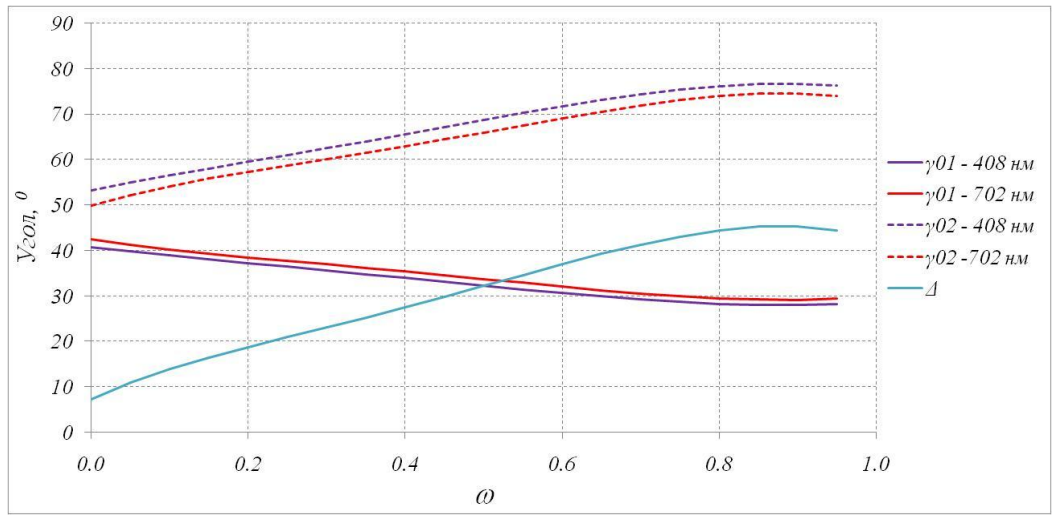

*Рис. 14. Зависимость параметров радуги от массовой доли серной кислоты*

При увеличении массовой доли серной кислоты от 0 до 90% для красного цвета угол $\gamma_{01}$ уменьшается в 0.7 раза от *42 36'* до *29 13'*, угол $\gamma_{02}$ возрастает в 1.5 раза от *49 57'* до *74 35'*. Угловой размер полосы Александра $\Delta = \gamma_{02} - \gamma_{01}$ возрастает более чем в шесть раз, с *7 20'* до *45 23'*.

Если измерять угловой размер полосы Александра в градусах, то аппроксимируя результаты расчетов полиномом, получаем

$$\Delta = D_0 + D_1\omega + D_2\omega^2 + D_3\omega^3 + D_4\omega^4,$$

где $D_0=7.4253$, $D_1=75.8766$, $D_2=-137.4143$, $D_3=242.6528$, $D_4=-146.1655$.

**Заключение**

В рамках геометрической оптики рассмотрен механизм формирования дождевой радуги. Приведены количественные оценки влияния концентрации серной кислоты в дождевых каплях на параметры радуги. Анализ углового распределения интенсивности света в радуге позволяет определить состав облака. Особенно чувствителен к изменению концентрации серной кислоты – угловой размер полосы Александра. Этот эффект может быть использован для дистанционной диагностики концентрации серной кислоты в облаках Венеры.